# Unraveling the chromophoric disorder of poly(3-hexylthiophene)

Alexander Thiessen[1], Jan Vogelsang[2], Takuji Adachi[2,3], Florian Steiner[2], David Vanden Bout[3] and John M. Lupton[1,2]

[1]Department of Physics and Astronomy, University of Utah, Salt Lake City, UT 84112, USA

[2]Institut für Experimentelle und Angewandte Physik, Universität Regensburg, D-93040 Regensburg, Germany

[3]University of Texas at Austin, Department of Biochemistry and Chemistry and Center for Nano and Molecular Science and Technology, 1 University Station A5300, Austin, Texas 78712-0165, USA

Classification: PHYSICAL SCIENCES: Chemistry

**_Statement of significance:_**

_Ideal photovoltaic cells would be black, absorbing all of the Sun's radiation, whereas Nature's machinery for solar energy harvesting – photosynthesis – looks green. Organic semiconductor devices, based on molecular building blocks, lie conceptionally between the extremes of inorganic and photosynthetic light harvesting. How can organic solar cells appear almost black if they are based on molecular units? Using single-molecule spectroscopy, we identify the fundamental electronic building blocks of_

*organic solar cells, and reveal that discrete molecule-like transitions scatter over the entire visible spectrum. The fundamental molecular unit is narrowband, but disorder induces a continuum reminiscent of that characterizing highly-ordered inorganic crystals.*

**Abstract:**

**The spectral breadth of conjugated polymers gives these materials a clear advantage over other molecular compounds for organic photovoltaic applications and is a key factor in recent efficiencies topping 10%. But why do excitonic transitions, which are inherently narrow, lead to absorption over such a broad range of wavelengths in the first place? Using single-molecule spectroscopy, we address this fundamental question in a model material, poly(3-hexylthiophene). Narrow zero-phonon lines from single chromophores are found to scatter over 200nm, an unprecedented inhomogeneous broadening which maps the ensemble. The giant red-shift between solution and bulk films arises from energy transfer to the lowest-energy chromophores in collapsed polymer chains which adopt a highly-ordered morphology. We propose that the extreme energetic disorder of chromophores is structural in origin. This structural disorder on the single-chromophore level may actually enable the high degree of polymer chain ordering found in bulk films: both structural order and disorder are crucial to materials physics in devices.**

Despite half a century of research into organic photovoltaics[1], the promise of versatile paint-on solar-cell modules based on conjugated polymers has prompted a present flurry of activity in the field[2-5]. A particular appeal of such excitonic solar cells is that very little material is needed to efficiently absorb light. This is due to the fact that the oscillator strength of primary photoexcitations, electron-hole pairs with binding energies far exceeding $kT$, is focused in the excitonic transition. The obvious downside is that this concentration also means excitonic transitions are inherently narrow and usually offer only mediocre spectral overlap with the broad solar spectrum. How then can excitonic solar cells be designed with appropriate spectral breadth? Merely introducing energetic disorder in the underlying excitonic material should lead to low-energy traps, impeding charge harvesting.

Although the optical and electronic properties of conjugated polymers are not perfectly suited to photovoltaics, their absorption spectra are surprisingly broad despite the excitonic nature of the transitions. Moreover, the diversity in functional characteristics revealed by varying processing conditions has fuelled the quest to formulate robust structure-property relationships between the electronic and optical properties and the underlying polymer structure[6-12]. Polythiophene derivatives have evolved into one of the chosen workhorse materials for solar-cell research[13,14]. Early spectroscopic studies established extraordinary solvato- and thermochromic characteristics, which were attributed to large conformational changes in response to the immediate environment[15-17]. It is little surprise then that such diversity in device characteristics exists when employing these materials[4]. Poly(3-hexylthiophene) (P3HT, structure shown in Fig. 1a) and related compounds were the first polymeric materials to exhibit signatures of two-dimensional electronic delocalization[18] due to a high level of interchain ordering most clearly visualized in x-ray diffractometry[19]. Despite this ordering, the optical absorption remains extremely broad, rendering the material favorable for photovoltaics. The striking dependence of P3HT ensemble optical properties on processing conditions has led to proposals that substantial dimerizing *interchain* interactions[20-23] could be involved. However, the signatures of excitonic dimerization, or H-aggregation, are not always straightforward to resolve conclusively[24]. H-aggregation should lead to a reduction in radiative rate[25]. Typically, upon aggregation of P3HT, a strong decrease in fluorescence quantum yield is observed but this is accompanied by an *increase* in fluorescence rate[26] since non-radiative decay rates are also enhanced. It is not always trivial to distinguish an increase in non-radiative decay from deceleration of radiative decay, complicating a precise determination of transition oscillator strengths. For this reason, most of the focus to date has been on interpreting spectral shape and vibronic coupling[23]. Although the origin of the magnitude of spectral shift between isolated and bulk-packed chains has been simply assigned to a "crystal shift"[23], a high degree of interchain ordering should also imply substantial planarization of the polymer with the associated increase in conjugation length[27] – which in turn decreases the strength of dipolar coupling that could lead to dimerization as the excited state becomes more delocalized[21].

Conclusively discriminating inter- from intrachain order is only possible by resorting to subensemble techniques such as single-molecule spectroscopy[28-35]. Here, we demonstrate the feasibility of reconciling fundamental spectroscopy of P3HT, shown in this study to be the most heterogeneous polymeric material studied to date on the single-chromophore level, with the excitonic single-chromophore picture[36-38] established for the most *ordered* polymers such as polyfluorenes[39] (PF), ladder-type poly(*para*-phenylenes)[40,41] (LPPP), polydiacetylene[42,43] (PDA) and poly(phenylene-ethynylene)[28] (PPE). On the level of single chromophores, P3HT behaves like other prominent materials, such as poly(phenylene-vinylenes)[41,44] (PPVs), with the exception of exhibiting giant variability in transition wavelengths spanning almost 200nm. The *dominant* photophysics of the ensemble – the striking red-shift between solution and film PL – is controlled by energy transfer to low-energy intrachain chromophores, in absence of the previously-proposed "crystal shift"[23].

## Results

### *Comparison of ensemble and single-chromophore spectroscopy*

Nominally different conjugated polymer materials exhibit very similar spectral characteristics on the single-chromophore level[28]. Since long polymer chains can contain many chromophores[40], it is not always straightforward to ensure that a single chromophore is identified within the polymer. Comparison with oligomeric model compounds, use of narrow-band excitation and cryogenic cooling of the sample has, however, enabled a reproducible framework for the identification of single chromophores[40]. We begin by comparing the ensemble photoluminescence (PL) spectra of P3HT in dilute toluene solution ($10^{-4}$ mM) and in a drop-cast bulk film in Fig. 1a. Going from solution to the solid phase shifts the spectrum by over 100nm (0.4eV) to the red and modifies the spectral form. The bulk spectrum has been interpreted to show a shift of oscillator strength from the 0-0 to the 0-1 transition, assigned to H-aggregation[22], since bulk-phase P3HT exhibits a high level of structural ordering in electron microscopy[5], scanning-probe microscopy[45] and other elastic scattering techniques[11,46,47]. The large spectral shift between solution and bulk has been assigned to the occurrence of a crystal shift in the H-aggregated chromophores embedded in an ordered environment of many other conjugated polymer chains[23]. Such a large spectral shift between solution and film is not observed in other common conjugated polymers. Six representative single-chromophore spectra, recorded at 4K from isolated chains directly deposited from toluene onto a $SiO_2$ surface, are superimposed on the ensemble spectrum in grey. The single-chromophore spectra all have the same shape, a comparatively narrow asymmetric zero-phonon line with a low-energy acoustic phonon wing and a distinct vibronic band offset by 180meV. The reduction of disorder broadening on the single-chromophore level facilitates determination of vibronic frequencies compared to the ensemble. The observed vibronics can be assigned to the ring C-C stretch (171meV) and the symmetric C=C stretch modes (179meV)[48]. The single-chromophore spectra, obtained under excitation at 458nm, span a spectral range of 195nm (0.8eV). In the red-most spectrum, excitation and emission are separated by 0.9eV, and yet the same universal[41] narrow single-chromophore spectral shape is observed.

To highlight the universality of chromophores, we plot 115 spectra in panel b, sorted by 0-0 peak energy (see Fig. S1 for alternative forms of presenting the data). Including the vibronic progression, narrow spectral lines are found between 485nm and 775nm. The general spectral characteristics, in particular the energy of the dominant vibronic plotted beneath, remain unchanged regardless of peak energy. The distribution of peak positions is plotted in a histogram in Fig. S2 and appears to be trimodal (see the discussion in the *Supporting Information* for possible origins of this distribution).

Interestingly, the 0-0 peak narrows slightly with decreasing energy, as plotted in panel c. Such spectral narrowing is seen in PF[28] and PDA[49], where lower-energy transitions correspond to improved chain ordering (the formation of the $\beta$-phase in PF). The opposite is seen in PPV[50], where chain bending induces a red-shift and an increase in spectral jitter. Over the entire range of chromophore energies, the ratio between 0-0 and 0-1 PL peak intensities scatters substantially but does not vary systematically with transition energy (panel d). Strong variations in the PL intensity of the vibronic progression between single chromophores have been reported previously, even in nominally rigid materials such as LPPP[51]. Such variations arise due to the slight distribution in ground-state molecular conformations, leading to differing structural relaxation energies and the resulting changes in the Franck-Condon progression, which become visible precisely because electron and vibronic transition are so narrow at low temperature. Examples of the interchromophoric scatter in 0-0/0-1 peak ratio are given in Fig. S3.

The single-chromophore PL spectra are identical in shape to that of different conjugated polymers[28] such as LPPP, PPV, PDA, PPE and PF as illustrated in Fig. S1, and exhibit the common blinking and spectral diffusion (spectral jitter, see Fig. S4). This similarity in spectral form strongly suggests that the narrow lines observed over such a broad spectral range arise from isolated chromophores[38]. Since narrow single-chromophore zero-phonon lines are observed as far to the red as 680nm, we conclude that the high-energy side of the bulk spectrum (red curve in Fig. 1) is appropriately described by the presence of isolated rather than aggregated chromophores: the red shift of over 100nm between solution and film can be mostly attributed to energy transfer from high-energy to low-energy chromophores. This conclusion, however, does not exclude the possibility of H-aggregate emission contributing to the red tail of the bulk PL spectrum. We stress that single chromophore spectroscopy, by its very definition, probes only the emission from single chromophores; we cannot probe potential H-aggregate emission without resorting to well-defined dimer structures[52].

*Difference in polymer chain conformation between solution and bulk phase*

Why do different chromophores dominate ensemble solution and bulk film spectra (Fig. 1a)? The fundamental difference between the two cases is found in the underlying chain conformation, which can be controlled through the polarity of the immediate environment of the polymer[30]: a "good" solvent, or matrix, will lead to optimal chain extension, driving the formation of well-solvated yet disordered spheres or globules. In contrast, a "poor" solvent promotes collapse of the polymer into toroidal or rod-like structures[29]. Solvent quality for non-polar organic compounds like conjugated polymers generally decreases with increasing polarity of the solvent. This effect can be replicated in the solid state for single chains of P3HT by embedding them in different polymer matrices. Figure 2a compares dilute ($10^{-4}$ mM, one hundred times the concentration employed in single-molecule experiments) solid solutions of P3HT in a virtually non-polar Zeonex derivative (Zeonex® 480) and in poly(methyl-methacrylate) (PMMA), which is more polar. The spectrum of P3HT embedded in Zeonex (green) closely resembles that of toluene solution, whereas the spectrum of P3HT embedded in PMMA (red) matches the peak and red tail of bulk P3HT. The mismatch in spectra at higher energy likely arises due to the presence of incompletely folded (i.e. blue-emitting) chains in PMMA. The effect of solvation is demonstrated in fluorescence micrographs of the films in Fig. 2b. The P3HT/Zeonex film appears uniform in emission whereas the P3HT/PMMA film shows discrete bright spots. Both images are displayed on the same intensity scale. In the P3HT/PMMA film, the background is darker than in P3HT/Zeonex, but the spots are much brighter. The formation of bright spots in PMMA suggests that multiple chains can aggregate together. In the following, we demonstrate that even *single* isolated chains in PMMA collapse into ordered structures.

Figure 3 reports measurements of the fluorescence modulation of single chains at room temperature under rotation of the polarization angle, $\theta$, of the exciting laser within the sample plane. For a straight object, the overall transition dipole should lie along the axis of the $\pi$-orbitals, leading to a strong cosine-squared modulation of PL with laser polarization[29]. Such a modulation in intensity is sketched in panel a. The depth of modulation, $M = (I_{max} - I_{min})/(I_{max} + I_{min})$, provides information on the degree of order regarding the transition dipoles of an individual chain. Examples of measurements are shown in Fig. S5. Fig. 3b compares histograms for 738 single chains in Zeonex and 587 chains in PMMA. Whereas the PL excitation (i.e. absorption) of the molecules in Zeonex is mostly weakly polarized due to disorder, P3HT in PMMA[31] is predominantly ordered with $M$ peaking around 0.8. Based on these $M$ values, possible chain conformations are sketched at the top of Fig. 4.

*Chromophoric emission at room temperature*

Intrachain conformation should have a dramatic impact on single-chain photophysics: in the random-coil structure, the chromophores will couple only weakly to each other, effectively emitting independently. In the ordered structures, energy transfer should occur between chromophores since interchromophoric distances are reduced. Two distinct experiments in Figure 4 clearly demonstrate this interplay between conformation and photophysics. The insets in Figure 4a show two representative single-molecule spectra at room temperature, measured in Zeonex and PMMA, which closely resemble the ensemble (Fig. 2a) for the solvated and collapsed structures, respectively. Panel a displays the PL intensity of a single chain as a function of time. The fluorescence beam is separated into two paths by a polarizing beam splitter and recorded with two separate photodiodes, allowing us to identify any change in orientation of the emissive transition dipole by quantifying the linear dichroism $D$ as defined in the schematic. An ensemble of different dipole orientations will lead to $D \approx 0$, as will a single dipole coincidentally oriented at 45° with respect to both detectors. The example P3HT chain in Zeonex is approximately five times brighter than in PMMA, but exhibits strong fluctuations and a gradual overall decrease in intensity (bleaching). The emission intensity does not drop completely to zero. In contrast, in PMMA, discrete blinking is observed. In Zeonex, the linear dichroism fluctuates around zero, exhibiting small jumps which imply the involvement of multiple different chromophore emitters. In contrast, in the PMMA example, a high static *D* is found: either all dipoles are oriented along the same axis or only one single chromophore in the polymer is active at once. The first conclusion is inferred from the excitation polarization modulation in Fig. 3b. Below, we demonstrate that indeed only one chromophore emits at a time from the single chain.

The number of multiple chromophores involved in the emission can be quantified by photon statistics obtained using a cross-correlation between photodiodes detecting the fluorescence divided into two pathways with a (non-polarizing) beam splitter (panel b)[53-55]. Fluorescence is excited by a pulsed laser (488nm) with a period of 25ns. If only one photon is emitted by the molecule per laser pulse, it cannot be simultaneously picked up by both detectors. Therefore, for single emitters, the cross-correlation must drop to zero at zero delay τ between detector signals, a phenomenon known as photon antibunching. Since such cross-correlation analysis requires high photon counts, we average[53] over 80 and 30 single molecules for PMMA and Zeonex, respectively. In Zeonex, the cross-correlation signal at τ=0 is nearly identical to that at other delays, implying that, on average, multiple chromophores on the polymer emit at once and do not couple efficiently. In PMMA, the cross-correlation at τ=0 drops to 20% of coincidence photon counts compared to τ≠0, implying predominant single-chromophore emission. Given the large number of chromophores on a chain, this phenomenon must result from energy transfer to the lowest-energy chromophore and concurrent singlet-singlet annihilation in the multichromophoric assembly[56].

**Discussion**

Cryogenic single-chain spectroscopy of P3HT reveals clear signatures of discrete intrachain chromophores. The diversity of spectral characteristics found in the ensemble when going from isolated to aggregated chains originates from these chromophores. Universal single-chromophore spectra, comparable in shape to those observed in many other materials, scatter over almost 200nm, providing a measure of the level of interchromophoric inhomogeneous broadening of ~0.8eV, greatly exceeding prior estimates[23]. In comparison, the corresponding disorder extracted from identical experiments for LPPP is 0.03eV and 0.2eV for PPV[41]. Such a level of energetic disorder as found in P3HT is *unprecedented* in molecular emitters, and there is no immediately obvious explanation for this scatter based on conventional models of conjugation in P3HT[57].

Ensemble P3HT shows a large spectral shift upon going from dilute solution to the bulk film (Fig. 1a). Since the transition between solution and film probes such a broad spectral range, intermediate states have been generated by means of solvent-vapour annealing in order to optimize both absorption and charge transport in solar cells[3,8]. Despite this heterogeneity of ensemble P3HT, key elements of the bulk photophysics[22-23] are consistent with the formation of discrete intrachain chromophores. We stress, however, that bulk spectral emission features extending beyond 700nm may be associated with H-aggregation or charge-transfer phenomena[1], which would only arise in the bulk and cannot be probed readily by single-molecule techniques. Notable examples of spectral characteristics of bulk PL which we cannot probe here are the monotonic changes of spectral shape with temperature[23], and the evolution of the vibronic progression and temporal red shift in time-resolved PL[58]. These features can both be modeled with the H-aggregate picture[23].

The same universal spectral features are observed independent of energetic separation between excitation and emission. Ultrafast dissipation of excitation energy in P3HT has previously been concluded from photon-echo spectroscopy and was attributed to efficient coupling to vibrations[59]. However, the dissipation of 0.9eV, the difference between excitation at 458nm and emission at 680nm, would require five quanta of the dominant vibration, which seems implausible. Since the emission intensity of all chromophores is comparable, we conclude that each chain most likely has very similar absorbing chromophores (presumably short conjugated units), which then populate the emissive chromophore by energy transfer. We note that when only one chromophore is present on a polymer chain, such as in $\beta$-phase PF, absorption and emission spectra exhibit the expected near-perfect mirror symmetry[39]. Even though multiple chromophores must exist on single P3HT chains, it is reasonable to assume that each single chromophore identified in Fig. 1 by its emission also has a complementary specific mirror-symmetric absorbing feature[39]. The coexistence of short and long chromophores within a single polymer chain, distinguished by their transition energy and vibronic coupling strength, has been demonstrated previously in polyindenofluorene using single-chain light-harvesting action spectroscopy[60].

Intuitively, one may be inclined to speculate that the most extended chromophores are also the lowest-energy units. However, slight bending of the backbone, chain torsion and changes in bond alternation can mask a strict correlation between conjugation length and lowering of the optical gap[50]. Semi-empirical calculations have suggested that even substantial bending of the $\pi$-system in polythiophene need not necessarily disrupt conjugation[57]. Given the high degree of ordering of single-chain P3HT seen in polarization fluorescence modulation (Fig. 3), it is tempting to surmise that in these objects the $\pi$-system is also the most extended[27]. However, the fact that the vibronic progression does not depend systematically on transition wavelength is crucial (Fig. 1d). In PDA, for example, the most extended chains show a dramatic *increase* of 0-0 transition oscillator strength relative to the 0-1 peak at low temperature, since the extended $\pi$-system constitutes an effective linear J-aggregate[43]. It is conceivable that exciton self-trapping limits excitonic coherence and thus the transfer of oscillator strength to the purely electronic (0-0) transition. Such an effect has been suggested for *β*-phase PF[39], which also shows substantial vibronic coupling in near-perfectly extended chains under cryogenic conditions, in contrast to PDA. It is also conceivable that the broad energetic distribution of zero-phonon lines (Fig. 1) arises from an intrinsic Stark shift due to trapped charges[61]. However, in this case a correlation should exist between increased red shift and increased linewidth[62], which is not observed; the linewidth appears to *decrease* with decreasing transition energy.

The most red-emitting single chromophores (Fig. 1) constitute the monomolecular precursor states for the red species in bulk P3HT films. Emission in the bulk occurs preferentially from these species, populated by energy transfer provided the polymer chain is sufficiently collapsed and ordered, as is the case in PMMA matrices or in the bulk. In this regard P3HT behaves like many other conjugated polymers, such as MEH-PPV[63-64], with the fundamental difference that interchromophoric energetic disorder is at least twice as large. As more and more chains pack together, the quantum yield of fluorescence of the mesoscopic objects within the bulk film is reduced since the probability of exciton dissociation and charge formation increases with greater average exciton migration distances[24,33]. While we have identified isolated chromophores which emit at wavelengths of ~680nm, a further red-shift with respect to this purely chromophoric transition observed in bulk films may arise due to either the previously described phenomenon of weak H-aggregation[22]; due to a Stark effect resulting from the electric field emanating from local trapped charges[65]; or from solid-state solvatochromic effects[1]. Finally, we note that it is both the conformation of the overall chain which changes with polarity (matrix) as well as the energetic distribution of chromophores which are responsible for such broad spectral variability. Nanoscale conformation therefore also affects ensemble absorption, enabling the tuning of absorption spectrum by solvent-vapour annealing[3,8].

The most obvious origin of the energetic spread in universal single-chromophore spectra lies in structural variations between single chromophores. We conclude that single chromophores are both flexible and can adopt a wide range of subtly-varying conformations: structural and energetic disorder dominates on the single-chromophore level. In contrast, in bulk films, an extraordinary high degree of order can be reached[19]. We propose that this unique feature of P3HT arises directly from the disorder on the single-chromophore level: as chains fold back on themselves or aggregate with other chains, there are always suitably-shaped chromophores present so that a closely-packed structure can be formed. It is conceivable that disorder on the single-chromophore level could actually breed order in the bulk.

P3HT constitutes one of the most heterogeneous material systems explored in the context of organic electronics, which, besides some of the impressive device characteristics[14], accounts for its continuing popularity in the field[13]. Nevertheless, conventional incremental materials development risks being impeded by lack of a robust understanding of primary photoexcitations in these systems. Is the primary photophysics of polythiophenes dominated by intermolecular interactions and H-aggregate species, or does the material actually adhere to the established concepts[36,38,41] of intrachain chromophores as derived from a wide range of compounds? Our single-molecule experiments tend to favor the latter notion, while leaving room for features of H-aggregation or other bulk solvation effects in the red-tail of the emission spectrum. In contrast to all polymeric materials studied previously, the energetic heterogeneity of chromophoric transitions spans almost 1eV, i.e. much of the visible spectrum. Such a breadth of possible fundamental transition energies can only be accounted for by an extreme sensitivity of electronic structure to conformational variations, which explains the dramatic impact processing conditions have on bulk material properties[3,8].

**Methods**

P3HT (poly(3-hexylthiophene), regioregularity = 95.7%, $M_w = 65.2kDa$, PDI=2.2) was purchased from EMD Chemicals and used as received. PMMA (poly(methyl-methacrylate), $M_w$=96.7kDa, $M_n$=47.7kDa) and Zeonex (Zeonex® 480) were obtained from Sigma-Aldrich Co. and Zeon Europe GmbH, respectively. Low-temperature single-molecule spectroscopy was carried out in a home-built fluorescence microscope as described previously[28]. The fluorescence was excited at 458nm using a frequency-doubled Ti:sapphire femtosecond laser system (HarmoniXX, APE GmbH and Chameleon Ultra II, Coherent Inc.). In order to minimize contamination of the weak fluorescence signal by background luminescence, single-chromophore spectra were recorded from samples deposited, without a polymer matrix, directly on top of $SiO_2$-covered Si wafers. The wafers were mounted in a liquid-helium cold-finger cryostat at 4K under vacuum. Concentration series were performed to ensure that the single-molecule density varied as expected with solution concentration. Due to the low photon count rates all experiments were carried out in spectral imaging mode (i.e. resolving the fluorescence spectrum[28]) rather than under direct two-dimensional imaging. This approach ensured that the observed fluorescence really did originate from single P3HT chains.

Room-temperature fluorescence was recorded on a separate microscope setup based on an Olympus IX71. P3HT chains were embedded in a poly(methyl-methacrylate) (PMMA) or Zeonex 480 host matrix according to a previously described procedure[59]: (i) Borosilicate glass cover slips were cleaned in a 2% Hellmanex III (Hellma Analytics) solution, followed by rinsing with MilliQ water. (ii) The glass cover slips were additionally bleached by a UV-ozone cleaner (Novascan, PSD Pro Series UV). (iii) The P3HT was diluted in toluene to single-molecule concentration ($\sim 10^{-12} - 10^{-13}$ M) and mixed with a 6% PMMA or Zeonex 480/toluene solution. (iv) This solution was dynamically spin-coated in a nitrogen glovebox at 2000 rpm onto the glass cover slips, leading to a film thickness of approximately 200 – 300 nm with an average P3HT chain density of ~20 individual P3HT chains in a range of $50 \times 50$ µm². The sample was incorporated into a home-built gas flow cell and purged with nitrogen to prevent bleaching by oxygen. Excitation was carried out by a fibre-coupled diode laser (PicoQuant, LDH-D-C-485) at 485nm under cw excitation for wide-field fluorescence microscopy or under pulsed excitation with a repetition rate of 40MHz for confocal fluorescence microscopy and time-correlated single-photon counting. The excitation light was passed through a clean-up filter (AHF Analysentechnik, z485/10), expanded and focused (or collimated) via a lens system onto the back-focal plane of a 1.35 NA oil immersion objective (Olympus, UPLSAPO 60XO) through the back port of the microscope and a dichroic mirror (AHF Analysentechnik, z488RDC) for wide-field or confocal excitation. For wide-field microscopy, the fluorescence signal was imaged on an EMCCD camera (Andor, iXon 897) after an additional magnification of $1.6\times$ and after passing a fluorescence filter (AHF Analysentechnik, RS488LP), whereas, for confocal measurements, the fluorescence signal was split either by a polarizing beam splitter (Thorlabs, CM1-PBS251) into two orthogonal polarizing detection channels or by a 50/50 beam splitter and detected by two avalanche photodiodes (APDs, PicoQuant, $\tau$-SPAD-20). The excitation intensities were set to ~1.5W/cm² and 150W/cm² for wide-field and confocal excitation, respectively. The polarization rotation of the excitation beam, confocal detection and photon statistics analysis were carried out as described previously[59]. For the antibunching measurements, we averaged the fluorescence traces of 80 single chains in PMMA and 30 single chains in Zeonex, respectively[51]. By considering the count rate, photodetector dark counts and the background intensity, we estimate the expected magnitude of the antibunching dip for a perfect single emitter as 10% and 2.5% for P3HT/PMMA and P3HT/Zeonex, respectively. These thresholds are indicated as dashed lines in Fig. 4b.

**Acknowledgements**

The authors thank D. Lidzey and G. Khalil for helpful discussions. JML is indebted to the David & Lucile Packard Foundation for providing a fellowship. AT acknowledges the Fonds der Chemischen Industrie for a Chemiefonds fellowship. This work was partially funded by the ERC Starting Grant MolMesON (#305020).

**Figure captions.**

Figure 1. **Unravelling the spectral heterogeneity of P3HT using low-temperature single-molecule spectroscopy.** a) Comparison of P3HT PL spectra in dilute toluene solution (green) and in a drop-cast bulk film (red) at room temperature. Six representative low-temperature (4K) single-molecule spectra (grey), consisting of a zero-phonon line and a vibronic progression, are shown as examples, spanning the spectral region from dilute solution to bulk film. b) Normalized PL spectra of 115 single molecules, sorted by the peak energy of the 0-0 transition. The same 0-1 vibronic transition, shifted by 180meV from the main peak is seen for all spectra (lower panel). c) Correlation of 0-0 peak width with transition energy. The green circles represent averages over 8 molecules, the red line is a guide to the eye. d) The intensity of the vibronic peak does not depend on chromophore transition energy.

Figure 2. **Replicating bulk and solution spectra in dilute matrix environments at room temperature.** a) The ensemble spectrum of chains dispersed in the inert Zeonex matrix (green) closely matches that of isolated chains in toluene (grey). In contrast, PMMA leads to chain aggregation: the spectrum of P3HT diluted in PMMA at the same concentration as in Zeonex (red) resembles the bulk film emission (grey). In both cases, the concentration of P3HT is one hundred times higher than in single-molecule experiments. b) The effect of matrix-induced phase separation is visible in fluorescence micrographs of the spin-coated films, plotted on the same intensity scale. In Zeonex, the film PL appears uniform, whereas in PMMA bright spots are seen corresponding to the formation of large multi-chain aggregates.

Figure 3. **Shape dependence of single P3HT chains on matrix material at room temperature.** a) The modulation of PL intensity under rotation of the plane of polarization of the exciting laser is recorded. The modulation depth $M$ provides a measure of chain extension and order: chains which mix well with the matrix (i.e. are well solvated) form random-coil structures which absorb any polarization of light. Poorly solvated chains, on the other hand, collapse, leading to ordered anisotropic rod-like structures. b) Histogram of $M$ values for single chains in Zeonex and PMMA. In Zeonex, isotropic structures are formed, whereas PMMA gives rise to highly anisotropic arrangements of the chains.

Figure 4. **Excited state properties of solvated (Zeonex) and collapsed (PMMA) single P3HT chains at room temperature.** In solvated chains, the chromophores emit independently of each other (cartoon). In collapsed chains, energy transfer occurs to the lowest-energy chromophore. a) Fluorescence intensity and linear dichroism $D$, determined by splitting the emission into two orthogonal polarization components as illustrated in the cartoon. In Zeonex, multistep blinking is observed. The emission is only weakly polarized ($D \approx 0$), exhibiting slight fluctuations as different chromophores switch on and off. In the PMMA collapsed-chain configuration, the fluorescence intensity is constant and shows single-step blinking. The emission is strongly polarized ($D \neq 0$) with no fluctuations. Representative single-chain PL spectra are shown in the inset. b) Photon

statistics in emission, given by the cross-correlation of two photodetectors in the emission pathway. The molecules are excited by laser pulses (25ns period), controlling photon arrival times $\tau$. In Zeonex, multiple chromophores emit, leading to only a 20% dip in photodetector coincidence rates. In PMMA, a dip in coincidence rate by 80% is observed, since no more than one single photon is emitted for each laser pulse. Based on photon counts and background signal for the two cases, the maximum dip expected for a perfect single-photon source is indicated by the dashed lines.

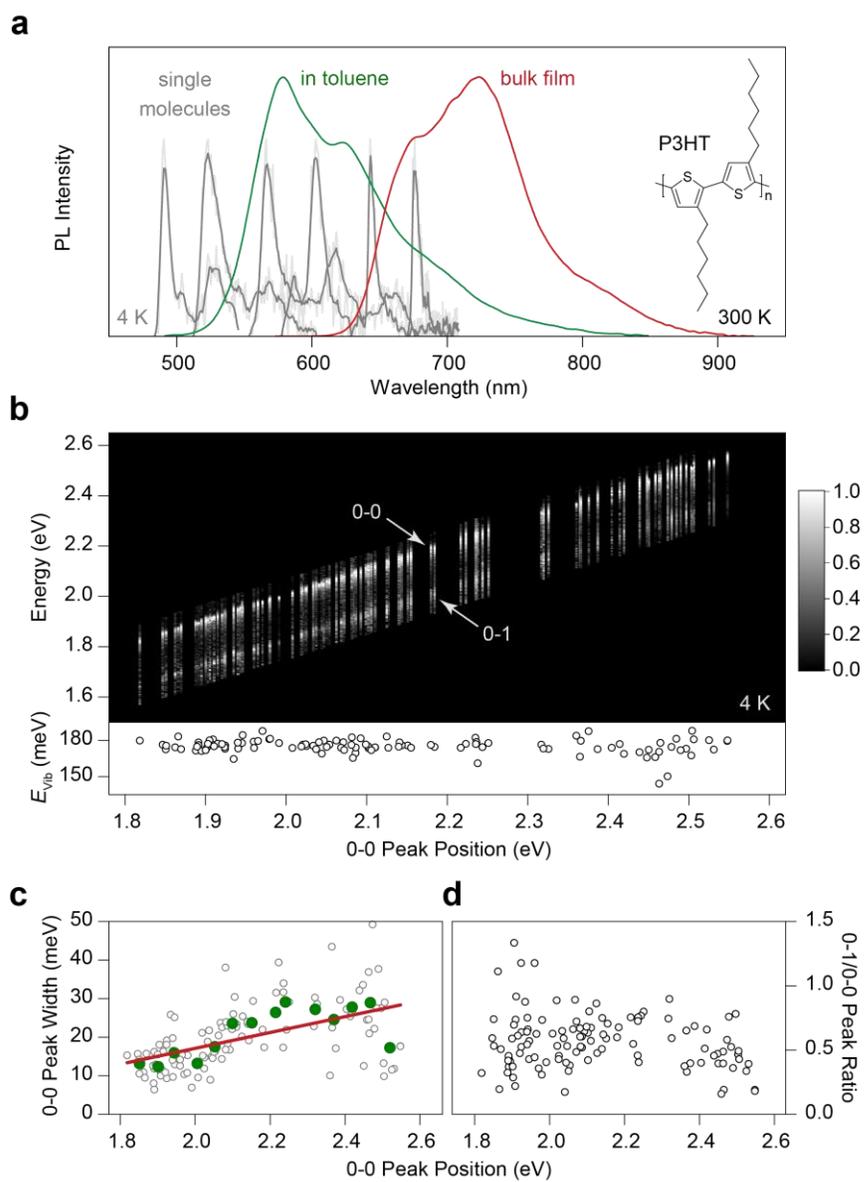

Figure 1.

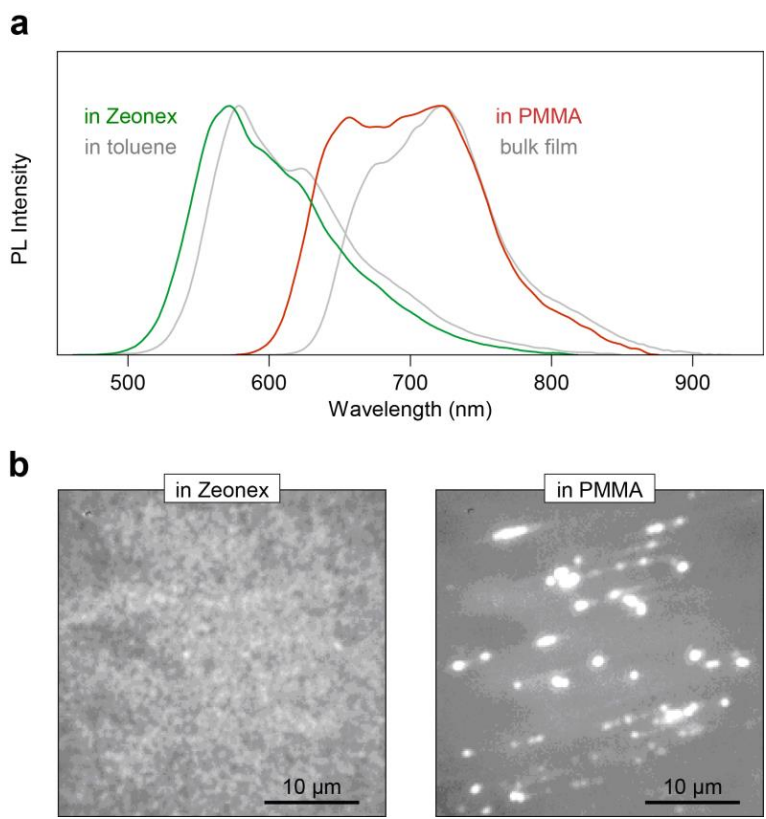

Figure 2.

**a**

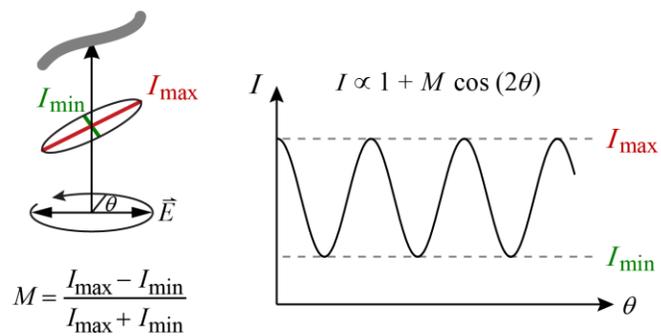

**b**

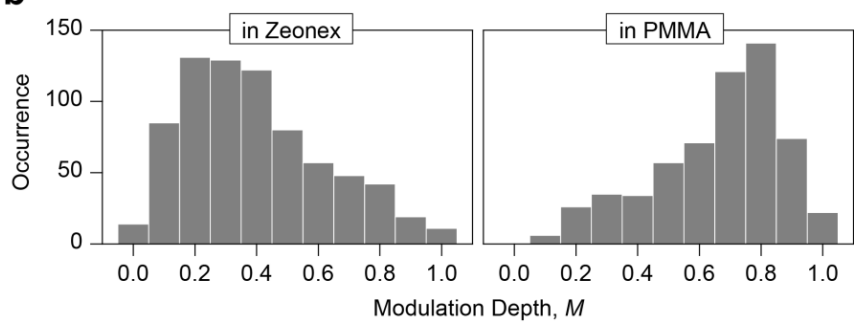

Figure 3.

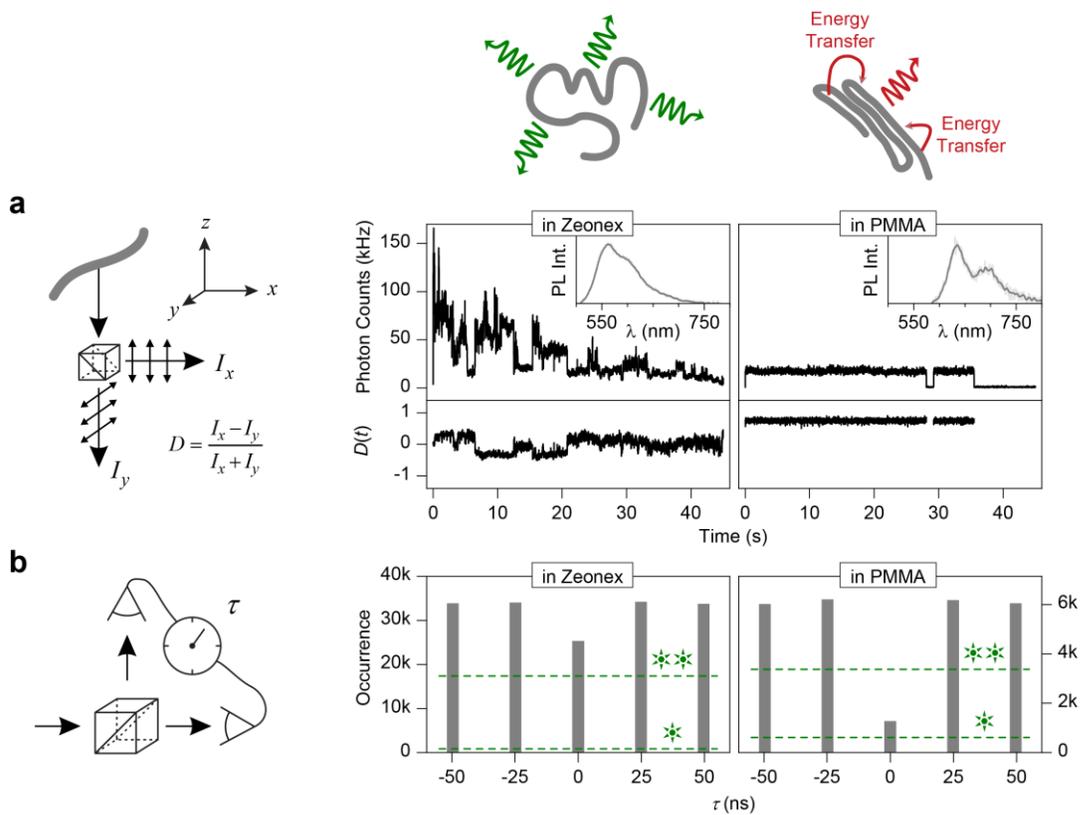

Figure 4.